# The effect of impurities on the mobility of single crystal pentacene

Oana D. Jurchescu, Jacob Baas, and Thomas T.M. Palstra

*Solid State Chemistry Laboratory, Materials Science Centre, University of Groningen,*

*Nijenborgh 4, 9747 AG Groningen, The Netherlands*

We have obtained a hole mobility for the organic conductor pentacene of $\mu = 35\ cm^2/V \cdot s$ at room temperature increasing to $\mu = 58\ cm^2/V \cdot s$ at 225 K. These high mobilities result from a purification process in which 6,13-pentacenequinone was removed by vacuum sublimation. The number of traps is reduced by two orders of magnitude compared with conventional methods. The temperature dependence of the mobility is consistent with the band model for electronic transport.

PACS numbers: 72.80.-r, 72.80.Le, 81.10. Bk.

Organic materials are presently being investigated and incorporated in semiconductor devices for a new era of the electronics industry. The understanding of the electrical conduction mechanism in these materials and at their interfaces represents a challenge, for which various, often conflicting models have been proposed. Organic conductors are conjugated materials where the π delocalized electrons are responsible for the intramolecular conduction. Molecular crystals are formed by relatively weak Van der Waals interaction between molecules, where the molecular packing determines the electronic behavior. Thus the charge carrier transport must be described using completely different models than for covalently bonded semiconductors. Of the many molecular conductors, pentacene is a promising candidate for future electronic devices and an interesting model system. Recent improvements in electronic aplications showed that this material exhibits mobilities higher than $1\ cm^2/V \cdot s$ for TFTs made from highly ordered films ([1], [2]). A mobility up to $8\ cm^2/V \cdot s$ was measured in single crystals of rubrene using a complete organic field-effect transistor [3]. The importance of impurities for the limitations in device performance has been emphasized during the last few years. However, little quantitative analysis concerning the consequences of impurities is incorporated in recent studies [4]. We report a mobility of $\mu=35 cm^2/V \cdot s$ at room temperature increasing to $\mu=58 cm^2/V \cdot s$ at 225 K for purified pentacene single crystals. The crystals were obtained by vapor transport growth in Argon flow after purification of the material by a vacuum sublimation technique designed to remove pentacenequinone. The content of the quinone impurity in pentacene was determined using high pressure liquid chromatography technique (HPLC), indicating a reduction by almost one order of magnitude.

The starting material for the experiment was pentacene obtained from Aldrich. Infrared absorption measurements (IR) (Nicolet Nexus) show that 6,13-pentacenequinone is present as impurity. The evidence for this is the presence of the absorption peak at *$1697 cm^{-1}$* which is assigned to a *$C = O$* bond vibration. The pure material does not have significant absorption in this region. The infrared experiments were complemented by mass spectrometry analysis, which confirmed that the $C = O$ vibration originates from a pentacenequinone molecule.

We have used *vacuum sublimation under a temperature gradient* as purification method. This technique is effective for separation of impurities from a solid if these impurities have a vapour pressure that is sufficiently different from the desired product [5]. The pentacene is placed in an alumina boat inside a glass tube that is thoroughly cleaned chemically, and then heated in a furnace under vacuum. The purification takes place at 430 K for 70 h under a dynamic vacuum of a membrane pump. Special attention is given to avoid contamination due to vacuum connections. The sublimated molecules will condense in athe cold zone of the tube. The entire set-up is in the dark to prevent UV degradation of the acene molecules. The carbonyl groups at each side of the middle ring reduce the sublimation enthalpy compared to the host molecule, thus at 430 K pentacene will not sublime and only quinone will be removed. This can be detected as a brown powder on the walls of the tube. The violet powder that did not sublime is purified pentacene that is used as the starting material for the single crystal growth.

Pentacene single crystals were obtained using physical vapour transport in a horizontal glass tube [6] under a stream of Argon. The use of ultra-pure Argon without Hydrogen as the transporting gas is motivated by the need to prevent the introduction of other impurities, like 6,13-dihydropentacene in the crystal, which can form by the hydrogenation of the acene at the middle ring (most reactive positions). The inner tube was cleaned by heating it under Argon gas flow to remove the solvents used for cleaning. The gas was obtained from AGA, with purity of 99.999%. A drying column that consists of aluminum oxide and hydrogen activated BTS-catalyst was inserted in the system for additional purification of the gas. We paid attention to the quality of the transporting gas because the quinone can be re-introduced by residual water or oxygen as ppm impurities in the carrier gas during growth. 30-40 mg of the source pre-treated material was placed at the end of the tube in an alumina crucible. A temperature gradient was applied by resistive heating of two heater coils around the tube. Without pre-purification of the pentacene, pentacenequinone will sublime together with the host molecule. Part of it will be introduced in the pentacene matrix in the crystallization process at the low temperature part of the tube. After obtaining ultra-pure crystals, they were annealed for 50 hours at a lower temperature (T = 450 K) than the crystallization temperature (T = 490 K). The crystals were stored in vacuum and darkness, as exposure to air and light will cause slow oxidation.

We used HPLC (Agilent 1100 LC/MSD) to determine the impurity concentration of the quinone. Pentacene and the pentacenequinone were separated on a silica column with a 3:1 v/v mixture of 1,2,4-trichlorobenzene and cyclohexane at 80°C. The amount of quinone was determined from the integrated intensity of the chromatogram using a diode array UV-Vis detector tuned at an absorption line ($\lambda$ = 390 nm). The quinone concentration was reduced from the as received material containing 0.68% in two sublimation steps to 0.17%. Subsequent crystal growth reduces the quinone concentration to 0.028% (±0.004) compared with 0.11% (±0.006) in crystals grown from untreated powder (figure1). The characteristic absorbtion of $C=O$ is observable even for the purest crystals (stage 5 in figure 1).

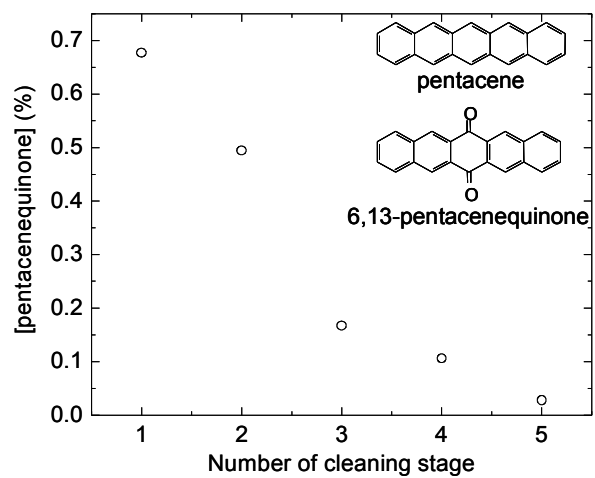

Fig1. 6,13-pentacenequinone concentration in pentacene in different stages of purification: 1-as received, 2-single sublimation clean, 3-double sublimation clean, 4-crystal grown from untreated powder, 5-crystal grown from doubly cleaned powder. The inset represents the Lewis structure of pentacene and 6,13-pentacenequinone.

We determined the electrical properties of the pure pentacene single crystals using space-charge-limited current (SCLC) measurements (figure 2). The samples were measured in darkness and a vacuum of $2·10^{-7}$ mbar. The mobility was calculated from the trap-free region ($\Theta=1$) of the space-charge-limited current regime, using the Mott-Gurney equation 1. We note that this formula was derived for the opposed electrode geometry, whereas we use a coplanar geometry.

$$J_{SCLC,tf} = \frac{9}{8}\varepsilon_0\varepsilon_r\Theta\mu\frac{V^2}{L^3} \quad (1)$$

where $J_{SCLC,tf}$ is the current density in the trap-free regime, $V$ is the applied voltage across a length $L$, $\varepsilon_r$ is the dielectric constant of the conductor (for which we use the literature value 3), and $\Theta$ the concentration of free carriers with respect to the total numbers of carriers (see equation 3).

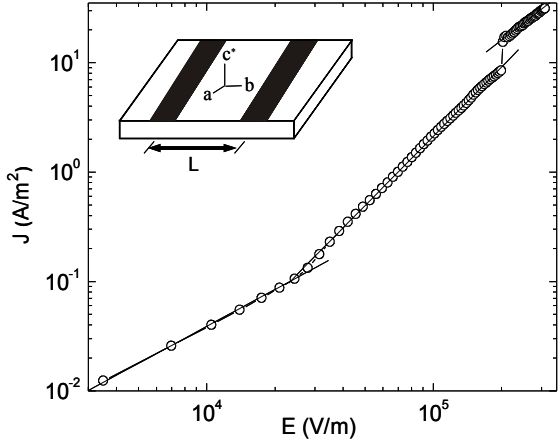

Fig 2. Current density vs. electric field for pentacene single crystal at room temperature. The solid lines represent the fits for the three different regimes. The inset shows experimental configuration of the a, b, and c*-axis, and the contacts. The hole-injecting electrodes are 40-50 nm thick gold.

The SCLC measurements show that the pentacene single crystals grown after pre-cleaning of the starting material are very pure, with $N_t = 1.74·10^{11}$ traps/cm$^3$. This is almost two orders of magnitude lower than the number of traps obtained for crystals grown with the conventional procedures [7].

If we assume a homogeneous current flow through the sample, the mobility is $\mu =11.2 cm^2/V·s$. However, as the mobility in the basal plane *ab* is much larger than the mobility along the *c\**-axis (perpendicular to *ab* plane), the current will be confined to the contact's side of the crystal. The measurements of the ohmic regime of the current-voltage characteristic showed different values for the resistivity for different directions ($\rho_a$= $1.3·10^6$ ohm·m, $\rho_b$= $4.7·10^5$ ohm·m, $\rho_{c*}$ = $2.1·10^8$ ohm·m). We have used Montgomery's method for analyzing anisotropic materials, transforming an anisotropic sample with resistivities $\rho_a$, $\rho_b$, and $\rho_{c*}$ and dimensions x, y, and z, to an isotropic solid with dimensions x', y' and z' [8, 9]. For the isotropic solid, the normalized effective thickness is determined to be $z_{eff}'/(x'·y') = 0.7$ for normalized sample thickness $z'/(x'·y')^{1/2} = 2.19$ and the ratio for the in-plane directions $y'/x' = 0.535$. Using this procedure, the relation between the effective thickness ($z_{eff}$) and the real thickness of the crystal (z) is calculated. Equation 2 expresses the conversion for the two dimensions:

$$z_{eff} = 0.32 \cdot z \quad (2)$$

With the effective thickness introduced in the current density $J_{SCLC,tf}$ in equation 1, the calculated mobility increases by more than a factor three (see figure 3). Therefore, a more accurate value for the mobility is $\mu =35 cm^2/V·s$ at 290 K and $\mu =58 cm^2/V·s$ at 225 K. We note that the Montgomery method is only valid in the linear part of the I-V regime. For I~$V^2$ the effective thickness ($z_{eff}$) should be considered as the upper limit. Thus, this analysis provides a lower limit of the intrinsic mobility.

Figure 3 shows that below room temperature the mobility increases with decreasing temperature following the relation $\mu = C·T^{-n}$ with $n=2.38$. We notice this behavior in several samples and it is consistent with a band model for charge transport in pentacene [10], with the interaction of the delocalized carriers with the phonons, the main scattering process. For electron-phonon coupling the mobility decreases with increasing temperature. This temperature dependence was also found in acenes with smaller number of benzene rings: naphthalene, anthracene, and tetracene [10 and the references there in]. Above room temperature a different transport mechanism dominates the mobility.

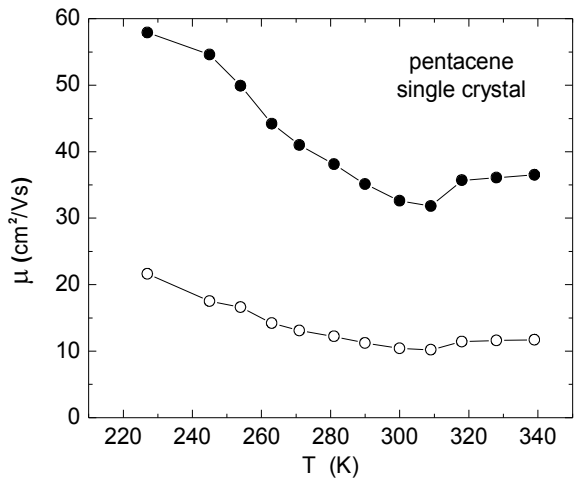

Fig 3. Temperature dependence of the electrical hole mobility for pentacene single crystal using the actual crystal thickness (○ open circles) and effective thickness (● closed circles).

In the first part of the SCLC regime (figure2), the injected carriers are trapped and the current is reduced by a factor $\Theta$, which represents the ratio between free and total number of charge carriers introduced in the solid (equation 3):

$$\Theta = \frac{n_f}{n_{tot}} = \frac{n_f}{n_f + n_t} \quad (3)$$

where $n_f$ and $n_t$ are the free and trapped carriers density, respectively, and $n_{tot}$ is the total carrier density. In our crystals $\Theta$ varies from 0.3 at 225 K to 0.82 at 340 K. With increasing electric field the density of injected carriers will increase, and above the trap-filled limit voltage ($V_{TFL}$) the mobility is not affected by impurity states and defects [11].

In the following we will focus on the origin of the trapping factor ($\Theta$) in equation 3. The traps in the crystal are mainly caused by structural imperfections and chemical impurities. Extended defects, such as edge dislocations or screw dislocations modify the available energy levels in their vicinity, often leading to the presence of accessible vacant orbitals in the band gap. We minimized the number of traps by a careful crystal growth and subsequent handling.

Heating the "as grown" crystals in an inert argon atmosphere will reduce the dislocations density. These defects are introduced during the crystallization process and are thermodynamically unstable. Thus their number decrease considerably by annealing. Dislocations will also enhance the chemical reactivity in their vicinity. Under the influence of light and temperature, reactions that oxidize pentacene to pentacenequinone will occur preferentially at dislocations [4]. So, even if the quinone is not present after crystal growth, it can be formed at defects after exposure to air and/or light.

We found that 6,13-pentacenequinone is the dominant chemical impurity. We didn't observe $C_{22}H_{15}$ and $C_{22}H_{13}O$ impurities that were calculated to form gap states in our pentacene, $C_{22}H_{14}$ [12]. Moreover, we argue that these molecules are irrelevant as these radicals are highly reactive. We were able to prevent the formation of the dihydropentacene $C_{22}H_{16}$ by using Argon as transport gas during the crystal growth. We have shown that the reduction of 6,13-pentacenequinone ($C_{22}H_{12}O_2$) impurities in pentacene by a factor five reduces the number of traps by almost two orders of magnitude. These impurities have different energy levels from pentacene, but they are energetically inert as a hole trap because their HOMO level is below that of the host molecule ([12] for $C_{22}H_{16}$ and [13] for $C_{22}H_{12}O_2$). This is distinctly different from experiments on smaller acenes, where the impurities yield stated in the gap [4]. For this reason, the number of traps in our measurements can be different from the number of chemical impurities. Although the impurity molecules do not act as trapping centers, they will induce a local deformation by distorting the pentacene lattice locally and create a scattering center. The quinone molecule is non planar and larger than pentacene. The middle ring has a flattened-chair shape with the *C=O* bond length of 1.216 Å and $\rangle C = O$ planar inclined at an angle of 3.1º to the molecular plane [14]. Thus, it will induce a local deformation leading to an increase in potential energy because of changes in molecular density. The quinone will strongly influence the number of such scattering sites, and thus the charge transport through organic single crystal.

In conclusion, we have reduced the impurity concentration of pentacenequinone in pentacene by a pretreatment consisting of vacuum sublimation of the impurity under a temperature gradient. The crystals exhibit a trap-free space charge current limited behavior. The mobility increases with decreasing temperature with a power law $\mu \sim T^{-n}$ from $\mu$ =35 $cm^2/V \cdot s$ at room temperature to $\mu$ =58$cm^2/V \cdot s$ at 225 K, indicating band transport. These results incorporate corrections for the effective thickness of the crystal for the anisotropic resistivity, where the effective thickness is at least three times smaller than the crystal thickness. Our results emphasize the importance of the control of defects and impurity states in molecular organic crystals in order to obtain a high electronic mobility, and allow studies of the band transport regime [15].

We would like to acknowledge P.W.M. Blom, C.R.van.den.Brom, A.F. England, R.A. de Groot, P. van't Hof, J.C. Hummelen, C. Tanase and G. de Wijs for experimental facilities and stimulating discussions. One of the authors (O.D. Jurchescu) would like to thank M. Mulder for constant encouragement, support and help.